\documentclass[cits]{PoS}

\title{Gravitational Radiation from Neutron Stars and Black Holes}

\ShortTitle{Gravitational Radiation from Neutron Stars and Black Holes}

\author{\speaker{Carlos F. Sopuerta}\\
        Institut de Ci\`encies de l'Espai (CSIC), Facultat de Ci\`encies, Campus UAB, Torre C5 parells, 
        E-08193 Bellaterra, Spain \\
        Institut d'Estudis Espacials de Catalunya (IEEC), Ed.~Nexus-201,  
        Gran Capit\`a 2-4, E-08034 Barcelona, Spain \\
        E-mail: \email{sopuerta@ieec.uab.es}}

\abstract{Gravitational Wave Astronomy is becoming a reality as Earth-based
interferometric gravitational-wave detectors reach the design sensitivities 
and move towards advanced configurations that may lead to gravitational-wave
detections in the immediate future.  In this contribution, I briefly summarize
the basic characteristics of this new area, the discovery prospects and the 
potential for fundamental physics.  Then, I present results of some investigations 
of two different sources of gravitational waves that are potential targets for 
present and future planned observatories.  First, I will discuss the generation 
of gravitational radiation by non-linear effects arising from the coupling between 
radial and non-radial oscillations of neutron stars, which may produce distinctive 
gravitational-wave signatures.  The gravitational radiation emitted by these 
sources is in the frequency band of Earth-based detectors.
And second, I will discuss the gravitational-wave emission during the inspiral
of extreme-mass-ratio compact binaries.  In this case, the gravitational waves
have low frequencies, inside the frequency band of space observatories like LISA.}

\FullConference{Supernovae: lights in the darkness 
                (XXIII Trobades Cient\'ifiques de la Mediterr\`ania)\\
		        October 3-5 2007\\
		        Ma\'o, Menorca, Spain}

\begin{document}

\section{Introduction}
Detection and study of the gravitational radiation coming from astrophysical
and cosmological sources is the main task of the emergent area of Gravitational Wave 
(GW) astronomy.  This task is coming closer to reality as Earth-based detectors
like LIGO~\cite{ligo} and VIRGO~\cite{virgo} are improving their sensitivities.  
At the same time, there is a strong international effort to 
construct a GW observatory in space, the 
{\em Laser Interferometer Space Antenna} (LISA)~\cite{lisa}, an 
ESA-NASA mission that is expected to be launched in the next decade,
with an ESA precursor mission, LISA PathFinder~\cite{pathfinder}, that will
be launched in 2010.

Gravitational radiation has physical properties that make it a very 
attractive tool for astronomy.  In particular, the weak coupling of
gravity with matter implies that the GWs that our
observatories will detect carry nearly uncorrupted information from 
the sources that produced them.  At the same time, this implies that GWs
are quite difficult to detect since they will also interact weakly with
the detector.  In many cases the signal will be buried
in the instrumental noise, and in some cases, in GW 
backgrounds (for instance, LISA will be limited by the GW background
produced by galactic stellar compact binaries).  Therefore, an important
task of GW astronomy is to construct theoretical waveform templates in order to separate
in an efficient way 
the signal from the noise using statistical techniques like match filtering.
Depending on the GW source, these templates can be obtained by 
full numerical relativity simulations, relativistic perturbation theory,
or post-Newtonian approximation schemes, or by a combination of them.
This is a research area in GW astronomy where a significant part of the theoretical
efforts have been directed to.  There are other important topics that theory is
dealing with that are important for the development of GW astronomy.  In particular,
the study of the formation mechanisms and rates of GW sources.  These studies
are relevant for the design of the GW observatories.  They can also determine
the regions of the configuration space of the potential GW sources that correspond 
to astrophysically realistic situations.  
This last point is crucial to establish the parameter space that GW template banks have to 
cover, and hence to determine the range of initial conditions for the simulations of GW sources.
For instance, in the case of black hole (BH) binaries we are interested in the
range of orbital parameters and spin orientations that we can expect from
each of the possible astrophysical scenarios that yield BH binaries. 
Finally, another important arena for theoretical studies is the investigation of 
possible electromagnetic counterparts to GW events (GRBs, X-rays, etc.) and their impact on 
on GW astronomy, whereby synergies between
electromagnetic and GW observations are explored.

The main sources of gravitational
radiation for Earth-based interferometric detectors are: stellar-mass compact 
binaries [BH-BH, BH-Neutron Star (NS), and NS-NS binaries], supernovae
core collapse, relativistic pulsars, oscillations of neutron stars (w-modes, r-modes, etc.),
and stochastic sources (astrophysical and cosmological). On the other hand, 
the main sources for space-based detectors like LISA are:
Galactic compact binaries [e.g. White Dwarf (WD) binaries], massive BH binary mergers,
capture of stellar-mass compact objects by massive BHs sitting at galactic centers
[known as extreme-mass-ratio inspirals (EMRIs)], and cosmological backgrounds. 
Remarkable advances in astrophysics, cosmology and gravitational physics are
expected from observations of these systems. 
To illustrate this, let us mention a few outcomes of GW astronomy: 
Direct proof of the existence of Gravitational Radiation
and its properties (we have an indirect proof of their existence through the observations
of the binary pulsar PSR B$1913+16$~\cite{binarypulsar} discovered by Hulse and
Taylor~\cite{hulsetaylor}); information about masses, 
spins, and orbital parameters for relativistic compact binaries;
tests of the {\em no-hair} theorem which, roughly speaking, says that astrophysical BHs 
are characterized only by their mass and intrinsic angular momentum;
information about the equation of state of NSs, 
rotational properties, progenitor masses, etc.; 
tests of galaxy formation models and understanding of the
growth history of massive BHs; study of the expansion history of the universe  by combining
gravitational and electromagnetic observations of sets of 
GW \emph{standard candles} like massive BH binary mergers or EMRI events; etc.
As a consequence, it is expected that GW astronomy will open a new window to the
exploration of the Universe in a similar way as it happened in the
past with the opening of new electromagnetic windows.   Moreover,
there is a considerable potential for observation of unexpected GW sources, and it is also
possible that these sources may provide information relevant not only for astrophysics
and cosmology but also for fundamental physics.

An outcome of GW astronomy that deserves a separate comment is the possibility of 
testing General Relativity and putting constraints on alternative
theories of gravity~\cite{cliffordwill}.  The first such constraints came from 
solar system  observations, but the gravitational fields involved in the solar system are
weak and the velocities are small as compared with the speed of light. 
Therefore, these observations can only test the non-radiative weak-field limit
of relativistic theories of gravity.  The situation changed after the
discovery of the binary pulsar PSR B1913+16~\cite{hulsetaylor} (which has been
followed by other binary pulsars: PSR B1534+12, PSR
J1141-6545, PSR J1829+2456, PSR J0737-3039), as this system involves regions with strong
gravitational fields (inside and outside the components of the binary).
The way in which certain alternative theories of gravity can be constrained
by binary pulsar observations is by the timing of phenomenological
\emph{post-Keplerian} parameters (see~\cite{damour} for a detailed and recent review
of tests of gravity theories by using binary pulsars).  For a given theory,
one can compute these post-Keplerian parameters (eight) in terms of the
Keplerian ones (five) and the masses of the binary (two).  Therefore,
the measurement of a post-Keplerian parameter determines a curve in the
mass plane (the plane of the possible masses of the binary components).  
Then, the measurement 
of three post-Keplerian parameters constitutes a test of the gravity theory under 
consideration.   Such a test is passed if the three curves intersect at one point 
(if one adds the measurement error bars the {\em thick} curves should just have a 
non-empty intersection).  In the cases in which $n$ ($n\geq 3$) post-Keplerian 
parameters are measured, $n-2$ independent tests of the gravitational theory 
can be performed.

However, the tests just mentioned obviously do not involve GW measurements. 
To illustrate the potential of GW astronomy in this respect, let us mention the constraints
that LISA observations can impose on the coupling parameter of
scalar-tensor theories of the Brans-Dicke type, $\omega^{}_{BD}$.
In~\cite{willyunes}, assuming a $10^3\,M^{}_\odot~BH\;+\;1\,NS$ inspiral at $50\;Mpc$  and $1$
year of integration prior the last stable orbit, it was shown that it is possible to get 
the following bound: $\omega^{}_{BD} > 3
\times 10^5$ (the Cassini mission~\cite{cassini} yields $\omega^{}_{BD} > 4\times 10^4$).
Moreover, it was also shown that LISA observations can also impose bounds on the Compton 
wavelength $\lambda^{}_g$ 
of the graviton~\cite{willyunes}:
Assuming a $10^6\,M^{}_\odot$ BH binary inspiral at $3\;Gpc$ we can get
$\lambda^{}_g > 5.4\times 10^{16}\;km$ (4 orders of magnitude larger than with solar system
dynamics).  In~\cite{lisaalternative}, these estimations were revised
by including the effect of spin couplings.  It was found that
the bound on the Brans-Dicke parameter is significantly reduced by spin-orbit and 
spin-spin couplings (factors of $10$ to $20$), while the bound on the graviton Compton 
wavelength is only marginally reduced (factors of $4$ to $5$). 

As it has been mentioned in this introduction, NSs and BHs (of different sizes) are
among the most important sources of GWs, mainly due to their strong gravity.  However,
this same property also makes them very challenging for the task of making theoretical
predictions of the shape and amplitude of the GWs that they emit.
In what follows, I will describe some work on the topic of the description of
sources of GWs in which I have participated with other researchers in the
recent past.  More specifically, I will discuss some results on the description
of two very different types of sources of GWs: (i) GWs from non-linear oscillations of
relativistic stars, which are relevant for Earth-based GW observatories.  
(ii) Simulations of EMRI events and the GWs emitted by them, which is of
interest for space-based observatories like LISA.

\section{Gravitational Radiation from non-linear oscillations of relativistic stars}
Neutron stars can undergo oscillating phases during their dynamical
evolution.  Astrophysical scenarios in which these oscillations can
be relevant for GW astronomy are: oscillations in a  binary
systems close to merger due to the tidal force exerted by the companion during the
coalescence;  oscillations in a proto-neutron star after the core bounce due
either to the bounce dynamics or to the fall-back accretion of
material, which has not been expelled by the supernova shock; 
etc (see~\cite{kostasnikos} for a recent review).

Gravitational radiation is one of the dissipative processes that damp
the stellar oscillations carrying away important information about the
physical properties and structure of the star.  
These oscillations generate high frequency GWs, above $500-600\;Hz\,$, 
and therefore are targets for Earth-based interferometers like LIGO and
VIRGO~\cite{ligo,virgo}.   In order
to increase the chances of a detection, it is
necessary to increase our theoretical understanding of the sources and
provide more accurate templates of the GW signals. 

Linear perturbation theory is appropriate to describe the dominant features
of stellar pulsations, but it neglects important details due to non-linear
effects.  For instance, we need non-linear perturbation theory in order to 
understand energy transfer between different oscillatory modes and how 
this can saturate the \emph{f-mode} and \emph{r-mode} instabilities of rotating stars
or limit the persistence of \emph{bar-mode} instabilities.
Other relevant non-linear effects are the development of \emph{g-mode} instabilities
in core collapse supernovae, and mass-shedding-induced damping of oscillations
in rapidly rotating stars.   On the other hand, there is a wealth of
literature dedicated to the understanding of the generation of
gravitational waves by linear oscillations of relativistic stars 
(see~\cite{andersson} and references therein).  However, we
know relatively little about non-linear dynamical effects (mainly due
to numerical studies. See, e.g.~\cite{dimmelmeier}).

Despite the fact that strong non-linear effects require a fully
non-perturbative approach, it is reasonable to expect that some 
interesting physical phenomena may involve a mild non-linearity
for which a second-order perturbative treatment should provide a reasonable 
description.  Along these lines, an interesting scenario to study 
is the one in which a NS is
oscillating radially and non-radially.  At first order, radial
oscillations of a spherical star don't emit {\it per se} any GWs, but they 
can drive and possibly amplify
non-radial oscillations and then produce gravitational radiation to a
significant level. For instance, radial and non-radial oscillations
are expected to be prevalently excited after a core bounce. Even
though the quadrupole component provides the dominant contribution to
the gravitational radiation, the radial pulsations may store a
considerable amount of kinetic energy and transfer a part of it to the
non-radial perturbations. As a result, this non-linear interaction
could produce a damping of the radial pulsations and hence, an interesting
GW signal. The strength of this signal depends naturally on
the efficiency of the coupling, which is the effect we want to
explore.  In addition, one may expect the appearance of
non-linear harmonics, which may also come out at lower frequencies
than the linear modes~\cite{dimmelmeier}, where
the Earth-based laser interferometers have a higher sensitivity.

Following this motivation, we started investigating whether non-radial oscillations
can be driven or even amplified through coupling by an internal radial
oscillation, regardless of the presence of an external source.  These
non-linear processes can occur, for instance, in a proto-neutron star
that is still pulsating.  A mainly radial pulsation could, for example,
drive the non-radial oscillations, either naturally present, or
excited through fall-back accretion.

To carry out this study we employed a non-linear
multi-parameter perturbative formalism~\cite{nonlinear1,nonlinear2}
in order to develop a gauge-invariant perturbative scheme for
studying, in the time domain, the coupling between the radial
pulsations and, both polar~\cite{nonlinear3} and
axial~\cite{nonlinear4}, non-radial oscillations.  The time-domain
numerical simulations in the axial case were done in~\cite{nonlinear4},
whereas in the polar case were done in~\cite{nonlinear5}.  For simplicity,
here we will focus on the axial case.

As it was shown in~\cite{nonlinear3,nonlinear4}, the structure of the
perturbative equations is hierarchical.  That is, first we need to
introduce a background spacetime that will represent the situation
corresponding to no oscillations, that is, a stationary star (equilibrium configuration).  
Then, we can compute the first-order
oscillations (radial and non-radial) of the stationary star, and finally, using all these
ingredients, we can compute the perturbations describing
the non-linear coupling between linear modes.

The equilibrium configuration (the background spacetime) is a perfect-fluid 
spherically symmetric relativistic star.  Such a model is determined by 
the solution of the Tolman-Oppenheimer-Volkoff 
equations given an equation of state.  In our case, we assumed a polytropic equation of state: 
$p = K\, \rho ^{\Gamma}$
with adiabatic index $\Gamma=2$ and $k = 100\;km{}^2$.  Prescribing a central mass
energy density $\rho_c = 3 \times 10^{15}\;g\,cm{}^{-3}$, one obtains the following
values for the stellar mass and radius: $M =1.26\, M_{\odot}$ and $R=8.862\;km\,$.

Then, we have two classes of linear oscillations: (i) the radial pulsations, 
which correspond to the $\ell=0$ mode, and (ii) the axial non-radial oscillations
with harmonic number $\ell\ge 2$.  The \emph{radial} perturbations are completely
described by a set of four perturbative variables, two metric perturbations and two
fluid variables, which obey three first-order evolution equations and
two constraints (reflecting the fact that there is just one single radial degree of 
freedom).  We can set up a hyperbolic-elliptic problem where the
Hamiltonian constraint is solved at any time step for one of the
metric variables.  The \emph{axial non-radial} perturbations can be described
by a system of two equations, the axial wave \emph{master} equation
and a conservation equation.  The former governs the evolution of the only gauge
invariant metric variable of the axial sector, let us call it $\Psi^{NR}$, 
while the latter is an equation for the axial velocity perturbation, $\beta^{NR}$.  
At first order, the stationary character of the axial velocity allowed us to study
separately the spacetime dynamical degree of freedom and its
stationary part.  In particular, the stationary solution describes the
differential rotation induced on the background star by the $\ell$-mode of the velocity 
perturbation, and the associated metric perturbation describes the 
dragging of inertial frames.

The \emph{axial non-linear coupling} perturbations can be again described by a gauge
invariant metric perturbation, let us call it~$\Psi^{C}$, and an axial velocity
perturbation, $\beta^{C}$. In this case, in the stellar interior, these non-linear
perturbations satisfy inhomogeneous linear equations.  The homogeneous part
of these linear equations is constructed exactly from the same linear operators
as the equations for the first-order axial non-radial perturbations.
The inhomogeneous part, the source term, is the sum of products of first-order radial 
perturbations times first-order axial non-radial perturbations. In the exterior we do 
not have matter fields, and hence the dynamics is described again by
Regge-Wheeler-type master equations.  Stellar interior and exterior communicate
through the general relativistic \emph{junction conditions} at the surface of the star.

We have evolved these perturbative equations for two types of initial
configurations: (i) a differentially rotating and radially pulsating
star, and (ii) the scattering of a gravitational wave by a radially
pulsating star.  We will focus on the former configuration, as the
dynamics of coupling oscillations is more interesting
(for a complete description see~\cite{passamonti}).

The initial configuration for the radial pulsations has been
excited by selecting specific radial eigenmodes. We have chosen an
origin of time such that the radial eigenmodes are described only by
the eigenfunctions associated with the radial velocity perturbation
$\gamma^{R}$ (see~\cite{nonlinear4}).  The eigenfunctions and eigenvectors
are found by solving the Sturm-Liouville problem associated with this
perturbative variable. Using a numerical code based on the relaxation method we 
were able to determine the eigenfrequencies of the radial modes with an accuracy  
better than $0.2$ percent, as compared with values in the literature. 
It has been checked that the simulations done with an arbitrary initial radial 
mode satisfy, with high accuracy, 
the Hamiltonian constraint and are stable for very long evolutions. 
The radial spectrum, which has been determined by applying a fast Fourier transformation
to  the time evolution, reproduce the results in the literature with an
accuracy better than $0.2$ percent. 

The perturbations describing the axial differential rotation are
obtained by 
expanding in vector harmonics the relativistic version of the \emph{j-constant rotation
law} and taking the first component, the one that is related to the
GW emission, that is, the $\ell=3$ mode.
The initial profile of the axial velocity perturbations contains two
specifiable parameters:  a parameter $A$ that controls the amount of
differential rotation, and the
angular velocity at the rotation axis $\Omega^{}_c$.  The value for $A$
has been chosen in order to have a smooth profile and a relatively
high degree of differential rotation, since for high $A$ the rotation
tends to be uniform and then the $\ell=3$ mode vanishes.  On the other hand,
we have chosen an angular velocity that corresponds to a $10\;ms$ rotation
period at the axis of the star.  Once we have solved the perturbative equations
and obtained the GW signal for this particular angular velocity, we can
make use of the linear character of the perturbative equations to obtain
the same result for any other angular velocity just by applying the
corresponding rescaling. 
Finally, in order to study the effects of the coupling between the linear
oscillations we have taken zero initial data for the coupling
perturbations.

\begin{figure}
\includegraphics[width=14cm]{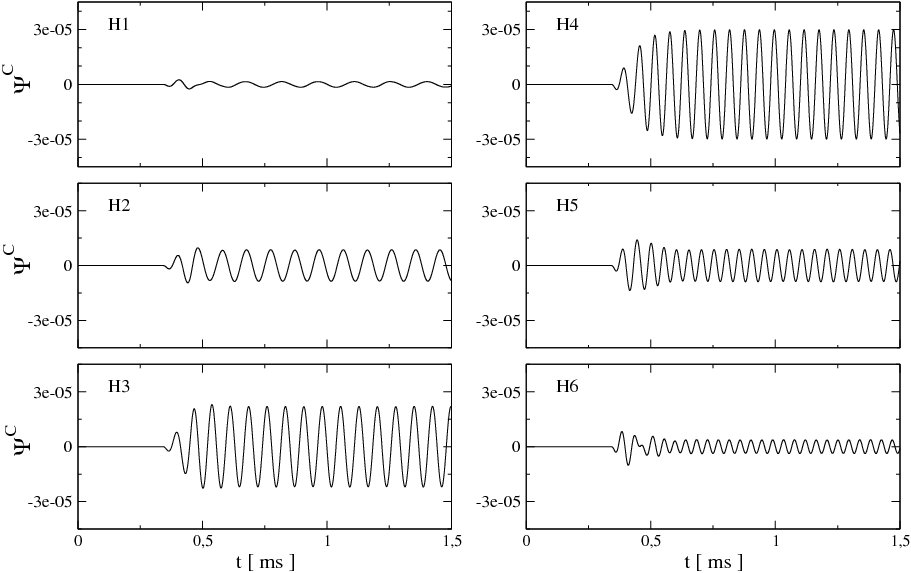}
\caption{\label{fig:Coupl_Psi11_H16} Comparison of six
$\Psi^{C}$ waveforms, in $km\,$, for the $\ell=3$ multipole.  The radial
pulsations considered correspond to single mode oscillation from H1 to
H6 overtone. These plots show that a resonance effect take place in
$\Psi^{C}$. See text for a discussion.}
\end{figure}

Numerical simulations of a radially pulsating and differentially
rotating star have shown a new interesting gravitational signal. The
gravitational waveforms exhibit the following properties: (i) An excitation of the first
$w$-mode at the beginning of the evolution. (ii) A periodic signal
driven by the radial pulsations present in the source term of the coupling
perturbations~(see Fig.~\ref{fig:Coupl_Psi11_H16}).  This picture is confirmed 
by the computation of spectrum of the perturbations, where we have noticed that 
the radial normal modes are precisely mirrored in the GW signal corresponding
to the non-linear perturbations.
On the other hand, the excitation of the $w$-mode at the early stages
of the numerical simulations is an unphysical response of the system
to small initial violations of the axial constraint equations for the
coupling perturbations.

In Fig.~\ref{fig:Coupl_Psi11_H16}, we can see that the metric waveforms 
associated with the coupling perturbations, and 
described by the axial master function $\Psi^{C}$,  exhibit an interesting amplification when
the radial oscillations pulsate at frequencies close to the $\ell=3$
axial spacetime $w$-mode, $\nu^{}_{w} = 16.092\;kHz\,$.
For the particular stellar model under consideration this effect appears at 
the third and fourth radial overtones, whose frequencies are $13.545\;kHz$ and 
$16.706\;kHz$ respectively.  It is worth mentioning 
that this effect takes place despite the fact that the energy and the maximum
displacement of the surface of the radial modes decrease
proportionally to the order of the radial modes~(see
Table~\ref{tab:damping_time}).  We can interpret this amplification
as a resonance between the radial frequencies, the source terms in the coupling
perturbative equations behave 
as forcing terms, and the natural frequencies of the axial oscillations, which
satisfy a wave-like master equation.

An important remark is that our perturbative approach does not include backreaction
effects.  That is, it cannot account for the damping of the radial oscillations
or the slowing down of the stellar rotation due to the contribution of the
non-linear coupling to the energy loss in GWs.
Backreaction effects can be studied by looking at higher perturbative
orders.  Nevertheless, we can provide a rough estimate of the damping
time of the radial pulsations by assuming that the energy emitted is
completely supplied by the first-order radial oscillations, and that
the power radiated in GWs is constant in time.  In
this way, the damping time is given by the following expression:
\begin{equation}
\tau^{C}_{\ell m} \equiv \frac{E^{R}_n}{<\dot{E}^{C}_{\ell m}>}\,,
\label{dam_time}
\end{equation}
where $E^{R}_n$ is the energy of a radial eigenmode~(see
Table~\ref{tab:damping_time}), and $<\dot{E}^{C}_{lm}>$ is the averaged
value of the non-linear coupling contribution to the power emitted.
The results for $\tau^{C}_{30}$ are shown in Table~\ref{tab:damping_time}.
In addition, in the last row of Table~\ref{tab:damping_time} we have given 
an estimation of the damping of the radial pulsations associated with a certain 
radial eigenmode in terms of the number of oscillation cycles:
\begin{equation}
N^{}_{\rm osc} = \frac{\tau^{C}_{\ell m}}{T^{}_{n}}\,,
\end{equation}
where $T^{}_{n} = \nu_{n}^{-1}\,,$ with $\nu^{}_{n}$ being the
eigenfrequency of the radial eigenmode. It is interesting to mention
that the number of oscillations required for the damping of the H4
mode is only $12$, and hence it would already affect the H4-waveform
shown in Figure~\ref{fig:Coupl_Psi11_H16}.  This is not surprising,
and shows that the coupling near resonances is a very effective
mechanism for extracting energy from the radial oscillations.

\begin{table}[t]
\begin{center}
\begin{tabular}{|c|c|c|c|c|c|c|}
\hline
        &              &              &           &       & & \\[-2mm]
 Radial &  Frequency   & $E^{R}_{n}$  & $\xi^{R}_{\mbox{sf}}$  & $<\dot E^{C}_{30}>$ 
& $\tau^{C}_{30}$  & $N^{}_{\rm osc}$ \\[2mm] 
 Mode   & $[kHz]$  & $~[10^{-8}\;km\,]$  &  $[m]$ &  $[10^{-14}]$ & $[ms]$ & \\[2mm] 
\hline
    F    & 2.138     & 35.9    &  12.65  &  $1.54\times 10^{-6}$   & $7.78\times 10^{10}$ & $1.67\times 10^{11}$ \\
    H1   & 6.862     & 4.2     &  4.02   &  $5.69\times 10^{-2}$   & $24.59\times 10^{4}$ & $1.69\times 10^{6}$  \\
    H2   & 10.302    & 1.37    &  2.66   &  $4.04$                 & $11.29\times 10^{2}$ & $1.16\times 10^{4}$  \\
    H3   & 13.545    & 0.62    &  2.02   &  $46.69$                & $44.28$              & $5.99\times 10^{2}$  \\
    H4   & 16.706    & 0.34    &  1.64   &  $130.84$               & $8.64$               & $1.44\times 10^{2}$  \\
    H5   & 19.823    & 0.21    &  1.38   &  $15.90$                & $44.04$              & $8.73\times 10^{2}$  \\
    H6   & 22.914    & 0.14    &  1.19   &  $3.71$                 & $126.12$             & $2.89\times 10^{3}$  \\
\hline
\end{tabular}
\caption{Quantities associated with
radial normal modes and their coupling to the first-order axial
differential rotation: Energy, $E^{R}_{n}$, and maximum stellar
surface displacement $\xi^{R}_{\rm sf}$ of the radial eigenmodes
for initial conditions with velocity
amplitude $0.001$; average power, $<\dot E^{C}_{30}>$, emitted in
gravitational waves to infinity from the coupling between the radial
eigenmode and the axial differential rotation; estimated values of the
damping times, $\tau^{C}_{30}\,;$ and number of oscillation
periods, $N^{}_{\mbox{osc}}\,,$ that takes for the non-linear
oscillations to radiate the total energy initially contained in the radial modes.}
\label{tab:damping_time}
\end{center}
\end{table}

In~\cite{nonlinear5}, a similar analysis has been carried out for the case of
polar non-radial oscillations, which have a richer spectrum than the axial ones.
For typical
values of mode-energies expected in the post-bounce phase of core-collapse
supernovae it has been found that some bilinear combinations of frequencies may
become detectable by the Advanced LIGO and VIRGO detectors~\cite{ligo,virgo}.  Detection
of these non-linear modes, 
together with the information coming from the expected linear modes, would provide new
constraints on the possible equations of state, as they also contain information on the
radial modes of the star.  

To conclude we would like to remark that these studies of non-linear oscillations of 
relativistic stars indicate the possibility of new interesting features in the GW 
signal emitted by these systems.  To explore further these ideas, additional work should
be done on the comparison with full non-linear simulations and also 
on improving the stellar model by including all the necessary physical ingredients
of astrophysically realistic stars.

\section{Gravitational Radiation from Extreme-Mass-Ratio Binaries}

Extreme-mass-ratio binaries in the inspiral phase of their evolution (EMRIs), i.e. when the system
evolves driven by GW emission,  
are considered to be a primary source of gravitational radiation to be detected 
by LISA~\cite{lisa}.   They consist of a stellar-mass compact object (SCO), such a main sequence star, 
a stellar-mass BH, or a NS (with masses in the range $1-10^2\,M^{}_\odot$ orbiting a massive BH (with 
masses in the range $10^5-10^7\,M^{}_\odot$), hence the 
mass ratios of interest lie in the range $10^{-3}-10^{-7}\,$.  
Their study is expected to provide crucial information about the growth history of 
galactic massive BHs, tests of the \emph{no-hair} theorem for BHs, tests of the validity of general relativity, 
possibility of discerning  among different theories of galaxy formation, etc.  

\begin{figure}
\begin{minipage}[c]{.6\linewidth}
\centering
\includegraphics[width=9.5cm]{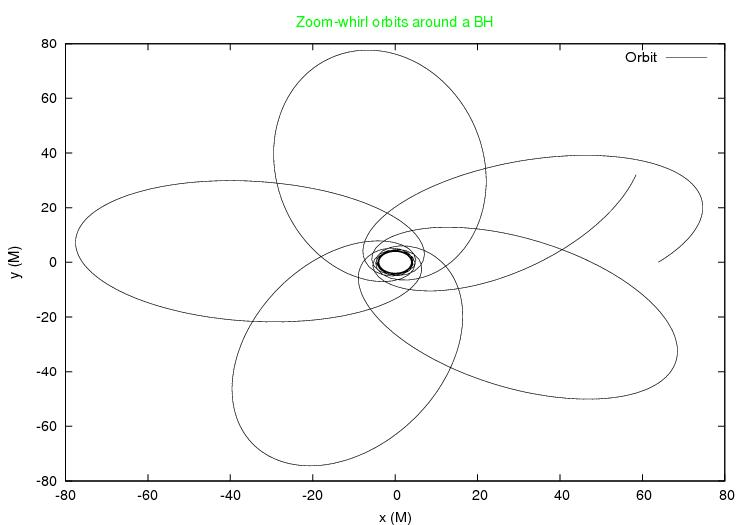}
\end{minipage}\hfill
\begin{minipage}[c]{.4\linewidth}
\centering
\includegraphics[width=5cm]{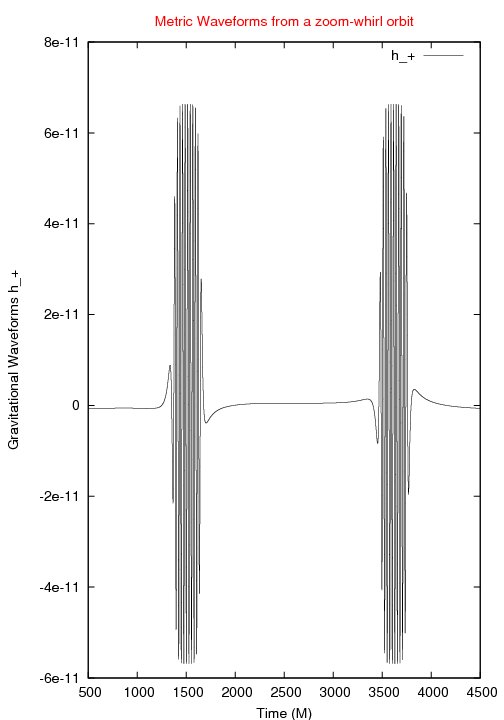}
\end{minipage}
\caption{\label{zoomwhirl} \emph{On the left:}~Zoom-whirl geodesic trajectory of a test mass around a non-rotating
BH.  The orbit can be parameterized as $r = pM/(1+e\cos\chi)$, where $M$ is the BH mass,
$p=7.8001$ is the semilatus rectum, and $e=0.9$ is the eccentricity. \emph{On the right:}~Gravitational
waveform (\emph{plus} polarization) associated with a portion of the zoom-whirl orbit.}
\end{figure}

Several astrophysical mechanisms that can lead to EMRI events have been proposed (for details,
see the review~\cite{paureview}).
The most studied is the one known as {\em single captures}, in which a SCO in the cusp
of the massive BH host galaxy is sent, due to 2-body or multi-body encounters, to an almost radial 
orbit towards the massive BH ($1-e\sim 10^{-3} - 10^{-6}$).  As the SCO approaches the massive BH, it loses energy
and angular momentum in GWs so that it gets bounded to the massive BH.  
The resulting orbit shrinks due to the GW emission and when the system enters the LISA frequency band
it still has a significantly high eccentricity ($e\sim 0.5-0.9$). The number of cycles during
the last year before plunge is of the order of $10^5$ cycles.  It has been 
estimated~\cite{gairetal,hopmanalexander}
that LISA may detect up to $10^3$ EMRI capture events per year.  In the last stages
of the inspiral, they SCO probes the strong-field region of the central massive BH and reaches velocities
very close to the speed of light.  To illustrate
this fact we present in Fig.~\ref{zoomwhirl} a trajectory that shows the so-called 
\emph{zoom-whirl} behaviour, characterized by strong precessional effects, and the
associated waveform, which gives an idea of the complex structure of EMRI waveforms.
Moreover, these EMRI events constitute precision
tools for GW astronomy since the GWs emitted carry a map of the massive BH spacetime (the massive BH
multipole moments) and it is expected that LISA will be able to measure 3-5 moments with high
precision~\cite{ryan,poisson,barackcutler}.  Moreover, it has been predicted~\cite{barackcutler} 
that the error in the estimations of mass, spin, and sky location from LISA observations are of the following order:
$ \Delta(\ln M^{}_\bullet)\,,~\Delta\left(\ln m/M^{}_\bullet\right)\,,~
\Delta\left(S^{}_\bullet/M^2_\bullet\right)~\sim~10^{-4}\,,~~
\Delta\Omega~\sim~10^{-3}\,.$

Other EMRI production mechanisms apart from single captures are: 
(i) Disruption of a stellar-mass compact binary passing near
a massive BH~\cite{emritipo2}. (ii) Capture of cores of giant stars close to the massive BH by tidal 
stresses~\cite{giantstars}. (iii) Inspiral of BHs produced in an accretion disc around the massive BH~\cite{bhaccretion}
(at distances $\sim 0.1-1\;pc\,$).  The EMRIs produced by this mechanism would have equatorial co-rotating 
circular orbits.  Actually, the different EMRI mechanisms mentioned here will produce distinctive GW signals that may
allow LISA to distinguish between them. 
Apart from these astrophysical events, it is interesting to consider two additional
systems that are closely related to EMRIs in the sense that can be described using essentially the
same techniques.  These systems are: (iv) Inspirals of intermediate-mass BHs, with masses in the range
$10^2-10^4\,M^{}_\odot\,$, located in globular clusters near the central massive BH~\cite{imbhinspirals}.
(v) Extreme-Mass-Ratio Bursts: long-period, nearly-radial orbits of compact objects around a central massive BH. 
These are quite similar to EMRI captures, the difference is that the compact object does not get captured,
and for this reason the GW emission consists of a few GW bursts produced in a few passages close to the massive BH.  
Therefore, only systems of galactic origin may be relevant for LISA~\cite{burstsyo}.

Since the GW signal from EMRI events will be buried in the LISA {\em noise}, it is crucial to 
have a good theoretical understanding of their evolution in order to produce accurate
waveform templates to be used in data analysis schemes for detection and extraction of physical
information from the LISA data stream.  Actually, it has been estimated that the theoretical waveform templates must
have a precision in the phase better than $1$ radian over $1$ week for detection, and better
than $1$ radian over $1$ year for extraction of physical parameters.

Due to the small mass ratios involved, BH perturbation theory is an adequate
tool for dealing with this problem.  The Kerr metric, describing a massive spinning BH,
will be the background spacetime to be used in the perturbative scheme, and it will obviously 
describe the central massive BH in the absence of other sources of gravitational field.  
The SCO can then be seen as a matter distribution that induces small spacetime perturbations.  
In order to compute these perturbations, one usually introduces the additional approximation of treating the SCO
as a point-like object.  This approximation is justified again by the small mass ratio, but
also by the fact that the internal structure of a compact object in General Relativity 
is {\em effaced} to a very high degree of approximation as post-Newtonian calculations have shown.  
More specifically, it has been estimated that the internal structure of a compact object contributes 
to order 5PN to the problem of motion.

A key ingredient for the accurate description of the motion of EMRIs is the
treatment of the backreaction of the perturbations, that is, how to evaluate the
effect of the perturbations generated by the SCO on its own trajectory.  
The equations of motion for the SCO including backreaction were consistently 
established in the works~\cite{misataquwa}, in the so-called {\em self-force} 
approach.  In this approach, the deviations from geodesic motion (around the BH background spacetime)
of the SCO are described in terms
of the action of a local force.  The practical implementation of the solution of
the equations of motion together with the determination of the gravitational waveform
is a difficult task that nowadays constitutes an active subject of research by various groups around
the world.  In part due to this fact,
several approximations to the description of EMRIs have been proposed.  In 
particular: (i) The {\em adiabatic} approximation.  It is based on the idea that the long-term
evolution may be approximated by the {\em dissipative} part of the gravitational
{\em self-force}~\cite{adiabaticaprox}.
(ii) {\em Klugde} waveforms.  The idea here is to use
approximations that can allow a quick generation of waveform templates
(in part for purposes of developing efficient data analysis schemes), which 
in practice means to find approximations that avoid solving partial differential equations~\cite{kludge}.

These approximations may provide useful templates for LISA detections of EMRI events
but it seems unlikely that they can provide templates for physical parameter estimations.
The key question in this respect is to what extend one can use adiabatic-type approximations, or in 
other words, how important is the conservative part of the self-force.
This is an issue that is presently under debate (see~\cite{debate}).  In any case,
self-force calculations can provide a method to compute precise gravitational waveforms 
for EMRI events, and at the same time,
these calculations can be used to test the validity of the adiabatic approximation.  

Broadly speaking, the study of the dynamics of EMRIs via the self-force approach involves a number
of challenges~\cite{generalmaterial} and can be divided into the
following three stages: (A) The computation of the gravitational perturbations
produced  by the SCO in the {\em background} spacetime of the massive BH, in particular
at the SCO location.  (B) Solving the equations of motion for the SCO including its
own gravitational field [computed in point (A)].  (C) Extraction of the relevant 
physical information, in particular the gravitational waveforms.   None of these parts has 
been completely solved yet although there has been a number of important advances in the last 
years.   Of particular relevance for the completion of this programme is the computation of the
perturbations and the evaluation of the self-force at the SCO location.  By one hand,
this requires numerical calculations since the equations involved cannot be completely
solved analytically and, on the other hand, the physical requirements on the precision
of the calculations are quite high.  Therefore, it seems that we need to have good enough
numerical schemes to perform these calculations.  In the last years, time-domain methods
have become more and more popular.  One of the main reasons is that they can 
produce accurate results for both very eccentric and almost circular orbits.  In contrast,
frequency-domain methods have the disadvantage that their convergence becomes very slow as we 
increase the eccentricity of the orbit.  

Numerical techniques in the time domain for self-force calculations is an active
area of research. Many of the developments are discussed in the annual \emph{CAPRA Ranch 
meetings on radiation reaction}.  The present author and collaborators have been
investigating the possibility of using Finite Element methods for this task.  First,
by studying a toy model in scalar gravity that contains all the ingredients (excepting
for the spin of the massive BH) of astrophysical EMRI events~\cite{emriyo1}.  The main conclusion
of this \emph{numerical experiment} was that the use of adaptive techniques for the description 
of EMRIs can significantly improve the efficiency and accuracy of the numerical computations.  
Second, Finite Element methods have also been used to compute the perturbations 
generated by a particle orbiting a non-rotating BH in the Regge-Wheeler gauge~\cite{emriyo2}.
In this work, it was shown how Finite Element methods can produce discretizations of
the Dirac delta distributions (and their derivatives) present in the perturbative
master equations and yield very accurate results for computations of GW fluxes of energy
and angular momentum for all type of orbits.  Future work will concentrate to extend 
this work to the case of computations in the Lorenz gauge, a gauge appropriate for
self-force calculations, and also to the case of rotating BHs.

\acknowledgments The author acknowledges support from the Ram\'on y Cajal Programme 
of the Ministry of Education and Science of Spain and by a Marie Curie
International Reintegration Grant (MIRG-CT-2007-205005/PHY) within the
7th European Community Framework Programme.


\end{document}